%% file: main.tex
\documentclass{llncs}
\newcommand{\norm}[1]{}
\usepackage{amsmath}

\usepackage[utf8]{inputenc}
\usepackage{graphicx}
\usepackage{listings}
\usepackage{amssymb}
\usepackage{dsfont}
\usepackage{mathtools}
\usepackage{algorithm}
\usepackage{algpseudocode}
\usepackage{xcolor}
\usepackage{multirow}
\usepackage{verbatim}
\usepackage{amsmath}
\usepackage[english]{babel}
\usepackage{wasysym}
\usepackage{qip}
\usepackage{qm}
\usepackage{enumitem}
\usepackage{environ}
\usepackage{booktabs}
\usepackage{pifont}
\usepackage[
	n,
	operators,
	advantage, 
	sets,
	adversary,
	landau,
	probability,
	notions,
	logic,
	ff,
	mm,
    primitives,
    events,
    complexity,
    asymptotics, 
    keys]{cryptocode}
\usepackage[normalem]{ulem}
\usepackage[colorlinks=true, linkcolor=black, citecolor=green]{hyperref}   
\usepackage[nameinlink,capitalise]{cleveref}

\spnewtheorem{Claim}{Claim}{\bfseries}{\itshape}
\usepackage[capitalise]{cleveref}
\input{commands}

\begin{document}
\title{Separating Pseudorandom Generators from Logarithmic Pseudorandom States}

\author{Mohammed Barhoush}
\institute{Universit\'e de Montr\'eal (DIRO), Montr\'eal, Canada\\  \email{mbarhoush18@gmail.com}}
\maketitle

\begin{abstract}
Pseudorandom generators ($\PRG$s) are a foundational primitive in classical cryptography, underpinning a wide range of constructions. In the quantum setting, pseudorandom quantum states ($\PRS$s) were proposed as a potentially weaker assumption that might serve as a substitute for $\PRG$s in cryptographic applications. Two primary size regimes of $\PRS$s have been studied: logarithmic-size and linear-size. Interestingly, logarithmic $\PRS$s have led to powerful cryptographic applications, such as digital signatures and quantum public-key encryption with tamper-resilient keys, that have not been realized from their linear counterparts. However, $\PRG$s have only been black-box separated from linear $\PRS$s, leaving open the fundamental question of whether $\PRG$s are also separated from logarithmic $\PRS$s.

In this work, we resolve this open problem. We establish a quantum black-box separation between (quantum-evaluable) $\PRG$s and $\PRS$s of either size regime. Specifically, we construct a unitary quantum oracle with inverse access relative to which no black-box construction of $\PRG$ from (logarithmic or linear) $\PRS$ exists. 

This does not directly separate $\PRG$ from some of the applications of $\SPRS$ since these applications involve, as a first step, a non-black-box construction of a notion termed $\bot$-$\PRG$s. To address this, we present another unitary separation showing that $\PRG$s are also separated from $\botPRG$s. Thus, we obtain separation from digital signatures and quantum public-key encryption.
\end{abstract}

\input{introduction_final}

\input{preliminaries_corrected_redline}


\input{first_separation_corrected_redline}
\input{second_separation_corrected_redline}
\bibliographystyle{splncs04}
\bibliography{mybib}

\end{document}

%% file: commands.tex
\newcommand{\hil}[1]{\ensuremath{\mathcal{#1}}}

\newcommand{\louis}[1]{}
\newcommand{\mo}[1]{}

\spnewtheorem{fact}{Fact}{\bfseries}{\itshape}
\crefname{fact}{Fact}{Facts}
\spnewtheorem{construct}{Construction}{\bfseries}{\itshape}
\spnewtheorem*{thm}{Theorem}{\bfseries}{\itshape}
\spnewtheorem*{lem}{Lemma}{\bfseries}{\itshape}

\newcommand{\PRS}{\ensuremath{\mathsf{PRS}}}

\newcommand{\AO}{\mathsf{O}_\ell}

\newcommand{\PRG}{\mathsf{PRG}}
\newcommand{\OWF}{\mathsf{OWF}}
\newcommand{\OWSG}{\mathsf{OWSG}}
\newcommand{\QPKE}{\mathsf{QPKE}}
\newcommand{\DS}{\mathsf{DS}}

\newcommand{\isbot}{\mathsf{Is}\text{-}\bot}

\newcommand{\SPRS}{\mathsf{SPRS}}

\newcommand{\LPRS}{\mathsf{LPRS}}

\newcommand{\botPRG}{\bot\text{-}\mathsf{PRG}}

%% file: introduction_final.tex

\begingroup
\emergencystretch=3em

\section{Introduction}
Pseudorandom quantum states ($\PRS$) were introduced by Ji, Liu, and Song~\cite{JLS18} as the quantum analogue of classical pseudorandom generators ($\PRG$)~\cite{BM84}.
Informally, an $n$-$\PRS$ generator is a quantum polynomial-time algorithm that, on input $k \in \{0,1\}^\lambda$, outputs an $n$-qubit state $\ket{\phi_k}$ that is computationally indistinguishable from a Haar-random state, even when an adversary is given polynomially many copies of the state.

Two size regimes of $\PRS$ have emerged as central: $(1)$ short pseudorandom states ($\SPRS$), where the output length $n$ is logarithmic in $\lambda$, and $(2)$ long pseudorandom states ($\LPRS$), where $n$ scales linearly with $\lambda$.
These regimes display markedly different behavior.  In contrast to $\PRG$s, no known reduction connects $\SPRS$ and $\LPRS$.  In fact, it was shown that there is no black-box construction of $\SPRS$ from $\LPRS$ \cite{CM24}, and there is evidence that the reverse direction is also difficult~\cite{BHM+25}.

{ Both $\SPRS$ and $\LPRS$ can be built from $\PRG$s~\cite{JLS18,BS20}.} In the other direction, Kretschmer~\cite{K21} proved a black-box separation between $\PRG$s and $\LPRS$s,\footnote{The separation is shown relative to quantum-evaluable one-way functions. 
{ Since standard deterministic $\PRG$s are also one-way, the result extends to those $\PRG$s as well.}} yet this does not rule out a reduction to $\SPRS$s. { His argument relies crucially on a concentration inequality for quantum states (Theorem 5.17 in \cite{M19}) whose bound is insufficient in the short-state regime.}  Our approach does not use this result, thereby circumventing that limitation.

The difference between the two regimes is also reflected in their cryptographic power. Long pseudorandom states enable several important primitives such as quantum pseudo-encryption, quantum bit-commitment protocols, length-restricted one-time signatures with quantum public keys, and private-key quantum money~\cite{JLS18,AQY22,MY22a}. Nevertheless, there remains a significant gap between the cryptographic power of $\LPRS$s and $\PRG$s.

Short pseudorandom states, in contrast, exhibit a much tighter connection to classical pseudorandomness: through quantum state tomography, one can extract a classical pseudorandom string from $\SPRS$s~\cite{ALY23}. {However, tomography is inherently probabilistic, leading to a notion termed \emph{pseudodeterministic $\PRG$s}---generators that are deterministic on most inputs but may behave non-deterministically on an inverse-polynomial fraction.} \footnote{Different notions of pseudodeterminism appear in the literature~\cite{B21,BBO+24}; we follow the definition of~\cite{ALY23}.} Nevertheless, this conversion  enables many of the applications of $\PRG$s, such as statistically binding commitments with classical communication. 

While pseudodeterministic $\PRG$s proved useful, their nondeterminism complicates their use as direct substitutes for standard $\PRG$s in some applications. To mitigate this, \cite{BBO+24} introduced the notion of \emph{$\bot$-pseudodeterministic pseudorandom generators ($\botPRG$s)}, {which they build from pseudodeterministic $\PRG$s by marking nondeterministic outcomes with the symbol $\bot$.} This seemingly modest modification provides a clean interface for cryptographic constructions to handle nondeterministic behavior explicitly. As a result, it enables $\SPRS$ to realize applications traditionally achievable only through standard $\PRG$s---such as many-time digital signatures and quantum public-key encryption with tamper-resilient keys~\cite{ALY23,BBO+24}---capabilities still beyond reach for $\LPRS$s.  

{Despite these advances, no complete quantum fully black-box (\textsf{BB}) separation is currently known between $\PRG$s and $\SPRS$s or even $\PRG$s and $\botPRG$s, leaving the precise relationship between them unresolved and somewhat limiting the significance of these applications.}

{ Resolving this question sheds light on the fundamental limits of quantum pseudorandomness and clarifies the relationship between these assumptions.}
{While partial progress has been made (see~\mbox{\cref{sec:related}}), a complete \textsf{BB} separation between $\PRG$s and $\SPRS$s or $\botPRG$s remains open.} The importance of this problem has been repeatedly emphasized in recent works~\cite{BBO+24,ALY23,BNY25}, underscoring its role in understanding the hierarchy of assumptions within the emerging framework of MicroCrypt quantum primitives.

\subsection{Our Results}

{In this work, we resolve this open problem by establishing two black-box separations.} Both separations hold relative to a unitary quantum oracle with inverse access, capturing a broad class of black-box constructions~\cite{CM24}. 

First, we separate $\PRG$s from $\ell$-$\PRS$s of any size $\ell\in O(\lambda)$, thereby also ruling out reductions to $\SPRS$.

\begin{theorem}{ There does not exist a \textsf{BB} construction of a $\PRG$ from an $\ell$-$\PRS$, for any positive integer-valued, efficiently computable function $\ell(\lambda)\in O(\lambda)$, even with inverse access.} \end{theorem}

We consider a general notion of $\PRG$s that is quantum-evaluable and permits a negligible probability of error or non-determinism. 

Moreover, prior works have shown that $\SPRS$s imply a rich class of cryptographic primitives. These applications proceed in three conceptual steps:
\begin{enumerate}
    \item Construct pseudodeterministic $\PRG$s from $\SPRS$ \cite{ALY23} \emph{(non-\textsf{BB})}. 
    \item Construct $\botPRG$ from pseudodeterministic $\PRG$s \cite{BBO+24} \emph{(\textsf{BB})}.
    \item{ Use $\botPRG$s to achieve some of the same applications achieved with $\PRG$ such as digital signatures \cite{ALY23,BBO+24}.} 
\end{enumerate}
{The first step, unusually, is a non-\textsf{BB} construction: the correctness of the pseudodeterministic $\PRG$ is guaranteed only by the \emph{security} of the underlying $\SPRS$.} Hence, the construction violates a requirement of \textsf{BB} reductions---namely, that the correctness of the target primitive must follow from the correctness of the source \cite{IR89,CM24}. In our separation, this property is used. Consequently, separating $\PRG$s from $\SPRS$s alone is insufficient to separate $\PRG$s from pseudodeterministic $\PRG$s and their many derived applications.

{ We therefore give a second separation that directly considers $\botPRG$s.} 
\begin{theorem}
\label{thm info 2}
    There does not exist a \textsf{BB} construction of a $\PRG$ from an $\botPRG$ (even with inverse access). 
\end{theorem}

{Previous work showed that $\botPRG$s can be used to derive many applications~\mbox{\cite{ALY23,BBO+24}}. As a result, we obtain the following as a direct corollary. }

\begin{corollary} \label{informal cor 1}{ There does not exist a \textsf{BB} construction of a $\PRG$ from any of the following primitives: }\begin{enumerate} 
\item {(Many-time) existentially unforgeable digital signatures for classical messages with classical keys and signatures ($\textsf{DS}$). 
}\item {CPA-secure quantum public-key encryption of classical messages with tamper-resilient keys and classical ciphertexts ($\textsf{QPKE}$). }\end{enumerate} \end{corollary}

{ We note that a version of \cref{thm info 2} was previously established by~\cite{BNY25} using completely positive trace-preserving (CPTP) maps, also ruling out constructions of one-way state generators (see \cref{sec:related}).} However, CPTP separations exclude a narrower class of constructions than unitary separations. In particular, a CPTP separation does not rule out reductions that rely on unitary access to the underlying primitive or adversary (see~\cite{CM24,BNY25} for further discussion of these distinctions). {In contrast, \mbox{\cref{thm info 2,informal cor 1}} establish, for the first time, general separations between these fundamental cryptographic primitives, encompassing reductions that may employ unitary or inverse access.}

\subsection{Related Works}
\label{sec:related}

Ananth, Lin, and Yuen~\cite{ALY23} were the first to formally investigate applications specific to $\SPRS$s. They showed how to construct pseudodeterministic $\PRG$s from $\SPRS$s, and used them to obtain applications such as statistically binding commitments with classical communication and CPA-secure symmetric encryption with classical ciphertexts.

A follow-up work~\cite{BBO+24} introduced the notion of $\botPRG$s, which they build from pseudodeterministic $\PRG$s, and use to achieve applications such as $\DS$ and $\QPKE$. 

{On the negative side, there are several separations that are relevant. Concurrent with this work, Bouaziz--Ermann, Hhan, Muguruza, and Vu \mbox{\cite{BHM+25}} introduced a novel conjecture on product states which, if true, would yield a separation between $\PRG$s and $\SPRS$s. Our result strengthens theirs by establishing the separation unconditionally, without relying on any unproven conjecture. Both separations employ a Haar random oracle. Their conjecture essentially asserts that if a $\PRG$ relative to this oracle is oracle-dependent---that is, outputs distinct values with non-negligible probability over the distribution of oracles---then the $\PRG$ must be non-deterministic on some oracles }\footnote{{An updated version of their work proves the conjecture, however this was done after our work.}}. Assuming this conjecture, they can directly rule out the existence of a deterministic $\PRG$, which is the hard part of our proof. We do not prove their conjecture, but establish a closely related guarantee via the continuity of quantum states, which suffices to carry out the separation and constitutes the technical core of this work.

Another closely related separation is due to Barhoush, Nishimaki, and Yamakawa \cite{BNY25}, who showed that $\botPRG$s are separated from one-way state generators ($\OWSG$s) relative to a non-unitary CPTP oracle. Recall that $\OWSG$s are the quantum analogue of one-way functions ($\OWF$s), and that every $\OWF$ is an $\OWSG$. By contrast, our separations are established relative to a unitary oracle. Notably, unitary separations rule out a broader class of constructions than CPTP separations. In particular, CPTP separations do not preclude reductions that rely on unitary or inverse oracle access to the underlying primitive or attack (see \cite{CM24,BNY25} for further discussion) and, hence, are considered weaker. At the same time, since $\OWSG$s are potentially a weaker primitive than $\PRG$s,{ their result separates $\botPRG$s from a weaker target, while our result allows stronger oracle access. These two results are therefore incomparable.} The reason a CPTP oracle admits a more direct separation in this setting is that it can encode non-determinism directly into the oracle, enabling the construction of inherently non-deterministic primitives such as $\botPRG$s while blocking purely deterministic ones.{ A unitary oracle must provide a coherent, reversible implementation, and allowing inverse access prevents a direct application of that argument.} {Our approach instead exploits the tension between the continuous nature of quantum states and the discrete nature of classical strings to rule out deterministic behavior.}

{All in all, both of the previous separation proofs \mbox{\cite{BHM+25,BNY25}} instill non-determinism property through either the oracle or through a conjecture.  Meanwhile, our techniques show that non-determinism is already inherent in procedures extracting classical randomness from quantum randomness.}

There are other notable separations in MicroCrypt, including the following:    
\begin{itemize}
    \item $(\LPRS \not\rightarrow \PRG):$ Kretschmer \cite{K21} presented a separation between $\PRG$s and $\LPRS$s, implying that $\LPRS$ constitute a potentially weaker assumption than $\PRG$s.
    \item $(\LPRS\not\rightarrow \textsf{DS}):$ Coladangelo and Mutreja \cite{CM24} showed that even digital signatures are separated from $\LPRS$.
    \item $(\LPRS\not\rightarrow \SPRS):$ This is a direct corollary to the separation above by~\cite{CM24}, but it was also shown through an alternative approach by \cite{BM24}. 
    \item $(\text{quantum-evaluable }\OWF\not\rightarrow \text{classical-evaluable }\OWF): $ 
{ Kretschmer, Qian, and Tal \cite{KQT24}} show a separation between classical-evaluable \textsf{OWF}s and quantum-evaluable \textsf{OWF}s relative to a classical oracle. As a corollary, they establish that classically-evaluable $\OWF$s are separated from $\DS$ and $\QPKE$. {However, our results imply that even quantum-evaluable $\OWF$s are separated from these applications.} Note that all notions in our work consider quantum-evaluable algorithms. 
\end{itemize}

\subsection{Limitations}

A primary limitation of our separation is that it focuses exclusively on \textsf{BB} constructions where the correctness of the target primitive is derived from the correctness of the underlying source. {This is a natural condition that is usually satisfied in cryptographic constructions.} In fact, the original definition of \textsf{BB} constructions \cite{IR89}, as well as its quantum adaptation \cite{CM24}, treats this ``correctness inheritance'' as a fundamental requirement. Furthermore, existing literature separating quantum primitives \cite{BNY25} similarly relies on this property, meaning those separations likewise do not account for constructions lacking this feature.

Nonetheless, one could envision a broader definition of \textsf{BB} constructions where target correctness may also stem from the security of the source primitive. Our current results do not exclude such cases, and extending our framework to encompass these ``looser'' constructions remains a compelling direction for future work.

That being said, this does not seem easy to do using our current proof techniques. Our proof relies on source correctness at multiple critical junctures. {This may be a natural consequence of the fact that any separation between $\PRG$ and $\botPRG$ needs to very delicately deal with the correctness of the source as this is the only factor separating these two primitives.}

\section{Technical Overview}

We now give an overview of our two separations. 

\subsection{Separating $\PRG$ from $\SPRS$}
We describe our separation between $\PRG$s and $\ell$-$\PRS$s for any function $\ell(\lambda)\in O(\lambda)$. The discussion below presents a high-level overview that omits several technical subtleties appearing in the full proof.

We use two unitary oracles in the separation:
\begin{enumerate}
    \item {{A \textsf{PSPACE} oracle $\mathcal{C}$, which is used to break any $\PRG$.}}
    \item{ {An $\ell$-Common Haar Function-Like State (\textsf{CHFS}) oracle \mbox{\cite{BHM+25}} $\mathcal{O}$, whose unitary implementation prepares a fixed state $\ket{\phi_x}$ on input $x\in\{0,1\}^*$.} The states are sampled independently from the Haar distribution and have size $\ell(\lvert x\rvert)$, where $\lvert x\rvert$ denotes the length of $x$.} 
\end{enumerate} 

{ The oracle $\mathcal{O}$ gives a $\PRS$ that is secure relative to $(\mathcal{O},\mathcal{C})$ with probability $1$ over the sampled states, by Theorem 4.4 in~\cite{BHM+25}.} 

Suppose, toward a contradiction, that there exists a \textsf{BB} construction of a $\PRG$ from a $\PRS$. This would imply the existence of a $\PRG$ $G^{\mathcal{O}}$ in this oracle model that may query $\mathcal{O}$. { {Because every choice of the states in $\mathcal{O}$ satisfies $\PRS$ correctness, $G$ must also satisfy correctness for every such oracle: the output is the same on every seed except with negligible probability.{ Note, however, that this negligible bound} may depend on $\mathcal{O}$.}}

{
We first obtain one correctness bound on a suitable family of oracles.
} Let $T=T(\lambda)\geq1$ be a polynomial bound on the number of queries made by $G$, and set $m=\lceil4\pi T\lambda\rceil+1$ and $\delta(\lambda)=1/(100\lambda m)$. Define the correctness error for $G$ when given access to oracle $\mathcal{O}$ as follows:
\begin{align*}
 e_{\mathcal{O}}(\lambda)
 &\coloneqq\mathbb{E}_{k\gets\{0,1\}^{\lambda}}
 \left[1-\max_y\Pr[G^{\mathcal{O}}(k)=y]\right]\end{align*}
Define the set of oracles with $\delta$ correctness error{ for all security parameters at least $N$} as follows:
 \begin{align*}
      F_N&\coloneqq\{\mathcal{O}:e_{\mathcal{O}}(\lambda)\leq\delta(\lambda)
 \text{ for all }\lambda\geq N\}.
\end{align*}
{
Correctness implies that every oracle belongs to some $F_N$.
} 

{
A \emph{prefix} fixes all states at labels $|x|<a$, for some constant $a$, and a \emph{completion} is an oracle that agrees with these fixed states. We show that some $F_N$ contains every completion of some prefix.

Suppose for contradiction that, no matter which finite prefix we fix and how large an integer $N$ we choose, there is a completion and a parameter $\lambda\geq N$ for which $e_{\mathcal{O}}(\lambda)>\delta(\lambda)$. This essentially gives us an increasing sequence of security parameters{ $\lambda_1,\lambda_2,\ldots$} where $e_{\mathcal{O}}(\lambda_i)>\delta(\lambda_i)$. We can carefully choose this sequence so that we first fix the behavior of the oracle up to $\lambda_1$, and then we search for the extension to $\lambda_2$ that satisfies the bound $e_{\mathcal{O}}(\lambda_2)>\delta(\lambda_2)$. Such a parameter exists by our assumption. 

We repeat this procedure, extending the prefix and choosing a larger parameter each time. Hence, this sequence is built inductively and defines one oracle for which the correctness bound fails at infinitely many parameters. This contradicts the correctness of $G$. Therefore, we get that some $F_N$ contains every completion of some prefix.
}

{
Define the oracle $I$ to have this prefix that keeps every completion in $F_N$ and let the oracle output $\ket{0^{\ell(|x|)}}$ at all other entries. Queries to $I$ can be approximated efficiently given a classical approximation of the prefix states. Note that the prefix is fixed, so these states have size that is constant in the security parameter, which allows us to efficiently approximate them. Thus, we can approximate oracle access to $I$.
}

{
For each oracle $\mathcal{O}$ with this fixed prefix, we construct a sequence $\mathcal{O}_1,\ldots,\mathcal{O}_m$ from $\mathcal{O}_1=\mathcal{O}$ to $\mathcal{O}_m=I$. Keeping the prefix unchanged, we move each full state vector, including its phase, along a shortest path on the unit sphere toward its value in $I$. Each path has length at most $\pi$. Dividing these paths into $m-1$ equal steps makes consecutive vectors close in Euclidean distance, so the corresponding swap unitaries satisfy
}
\begin{align}
\label{eq:trace}
 \|\mathcal{O}_i-\mathcal{O}_{i+1}\|_{\mathrm{op}}
 \leq\frac{4\pi}{m-1}.
\end{align}
Note that in this hybrid argument, we use the continuity property of quantum states, which does not hold for classical strings. In particular, we use continuity to{ make these small modifications} to the states in an undetectable manner from the perspective of the generator. $G$ queries the oracle at most $T$ times, so we get that for every seed $k$ and $i\in[m-1]$,
\begin{align}
\label{eq:trace-dis 1}
 \textsf{Tr}\left(G^{\mathcal{O}_i}(k),G^{\mathcal{O}_{i+1}}(k)\right)
 \leq\frac{4\pi T}{m-1}\leq\frac{1}{\lambda}.
\end{align}
This also handles inverse queries, since the swap oracles are self-inverse.

{  {For any seed, the output of $G$ cannot change between{ consecutive oracles: $G$ is deterministic, so} two different outputs would contradict~\mbox{\cref{eq:trace-dis 1}}. Thus $G^{\mathcal{O}}$ and $G^I$ have the same output with high probability on almost all seeds.}

{Since $I$ is efficiently simulatable, a \textsf{PSPACE} computation can estimate the output probabilities of $G^I$. This gives a constant distinguishing advantage, contradicting the security of the $\PRG$.}}

\subsection{Separating $\PRG$ from $\bot$-$\PRG$}

We now describe how to separate $\PRG$ from $\botPRG$ using a unitary quantum oracle with inverse access. In particular, we use two oracles:
\begin{enumerate}
    \item {{A \textsf{PSPACE} oracle, used to break any $\PRG$.}}
    \item  A modified quantum random oracle with an abort mechanism $\mathcal{O}$, designed to satisfy the pseudodeterminism property of $\bot$-$\PRG$s.{ {At each input length $d$, an inverse-polynomial fraction $\mu(d)$ of the inputs may produce either a fixed string or the symbol $\bot$; all remaining inputs produce a fixed string. }}
\end{enumerate}
{{It is straightforward to show that $\mathcal{O}$ is a valid $\bot$-$\PRG$.}} Suppose, toward a contradiction, that there exists a \textsf{BB} construction of a $\PRG$ from a $\botPRG$. This would imply the existence of a $\PRG$ $G^\mathcal{O}$.


{ The first step is similar to that of the previous proof.} We establish for the oracle $\mathcal{O}$ the existence of{ an integer $N$ and a prefix} such that every completion of the prefix is in $F_N$. 

The issue is the second step of the previous proof where we applied a hybrid argument using the continuity of quantum states to show that for any oracle with some prefix, the $\PRG$ can be simulated efficiently without oracle access. However, for this separation, we cannot use the continuity of quantum states since the oracle no longer outputs quantum states.

Instead, we apply a similar argument using the non-determinism of the oracle $\mathcal{O}$. In particular, $\mathcal{O}$ has{ an inverse-polynomial fraction $\mu$ of inputs} that are allowed to be non-deterministic. On such inputs, the output can be $\bot$ with {any  arbitrary probability}. Hence, we can utilize the continuity of this $\bot$ probability in place of the continuity of quantum states.{ Note, however, that we cannot} simply vary the probability of $\bot$ on all inputs, since only a $\mu$ fraction of inputs are allowed to output $\bot$. 

More specifically, for each non-deterministic input, we gradually increase the abort probability to $1$, then change its fixed string to $0$ while it always aborts, and then gradually decrease the abort probability to $0$. The unitary implementation of the oracle varies continuously along these changes. This is because we can change the non-$\bot$ output to any value when the probability of outputting $\bot$ is 1 without changing the output of the oracle. These modifications have converted the non-deterministic inputs to return $0$ with probability 1. 

Now the oracle is completely deterministic, so it gives us the ability to choose{ another $\mu$ fraction of inputs} and start increasing the probability  of outputting $\bot$ until it is 1, then change the non-$\bot$ output to $0$, and then gradually decrease the probability of $\bot${ to $0$.}

We repeat this process for all blocks at each length that $G$ may query. There are only polynomially many relevant blocks because $\mu$ is inverse-polynomial. Subdividing each change finely enough therefore gives the required polynomial-length sequence of hybrids that allows us to show that $G^{\mathcal{O}}$ can be simulated efficiently with some degree of accuracy. 

{ {
The \textsf{PSPACE} test then distinguishes the output of $G^{\mathcal{O}}$ from uniform with constant advantage. This contradicts the assumed \textsf{BB} construction.
}}

\subsubsection{Acknowledgments:} We would like to thank Samuel Bouaziz--Ermann for helpful discussions and an anonymous reviewer for finding a small issue in a previous version of this work, which we have corrected.{ We also acknowledge the use of ChatGPT Astra specifically for this version of the work for helping us simplify the fixed-prefix argument and obtain a shorter proof. 
} 
\par\endgroup

%% file: preliminaries_corrected_redline.tex

\begingroup
\emergencystretch=3em

\section{Preliminaries}

\subsection{Notation}
{ We let $[n]= \{1,2,\ldots,n\}$ and let $\negl[x]$ denote any nonnegative function that is asymptotically smaller than the inverse of any positive polynomial. All logarithms are base $2$.} 

{ We let $x\leftarrow X$ denote that $x$ is sampled according to the distribution $X$.} If $X$ is a set, then $x\leftarrow X$ simply means $x$ is chosen uniformly at random from the set. We let $\Pi_{m,n}=(\{0,1\}^n)^{\{0,1\}^m}$ denote the set of functions mapping $\{0,1\}^m\rightarrow \{0,1\}^n$. 

We refer the reader to \cite{NC00} for a detailed introduction to quantum information.{ We let $\mathcal{S}(\hil{H})$ and $\mathcal{U}(\hil{H})$ denote the sets of unit vectors and unitary operators, respectively, on the Hilbert space $\hil{H}$. We let $\textsf{Haar}(\mathbb{C}^d)$ denote the uniform probability measure on $\mathcal{S}(\mathbb{C}^d)$.}{ We let $\textsf{Tr}(\rho,\sigma)=\frac12\|\rho-\sigma\|_1$ denote the trace distance between density matrices. For classical distributions, this is the total variation distance, $\textsf{Tr}(p,q)=\frac12\sum_x|p(x)-q(x)|$.}

{ We follow the standard notation for quantum algorithms.} We say that a quantum algorithm $A$ is \emph{QPT} if it is a uniform polynomial-time quantum algorithm.

{ The security parameter $1^\lambda$ is implicit in algorithm inputs when omitted. }{ For an algorithm $A$ with classical output and a family of inputs $x=x(\lambda)$, we say that $A(x)$ is \emph{non-deterministic} if two independent evaluations yield distinct values with non-negligible probability in $\lambda$. At a fixed parameter, we say that $A(x)$ is \emph{$c$-non-deterministic}, for $c\in(0,1)$, if this probability is at least $c$.} 

\subsection{Pseudorandom Primitives}

We define pseudorandom states ($\PRS$s), first introduced in \cite{JLS18}.

\begin{definition}[Pseudorandom State Generator]
\label{def:prs}
    {Let $\lambda\in \mathbb{N}$ be the security parameter and let $n=n(\lambda)$ be a function in $\lambda$. A QPT algorithm $\textsf{PRS}$ is called a $n$-\emph{pseudorandom state generator (\textsf{PRS})} if the following holds:}
    \begin{itemize}
        \item \emph{(Correctness)}{ On any input $k\in \{0,1\}^\lambda$, $\textsf{PRS}(k)$ outputs an $n$-qubit state.}
        \item \emph{(Security)} For any polynomial $t(\cdot)$ and QPT distinguisher $\adv$:
        \begin{align*}
            &\left|\Pr_{{k}\leftarrow\{0,1\}^\lambda}
            \left[\adv\left(\textsf{PRS}({k})^{\otimes t(\lambda)}\right)=1\right]\right.\\
            &\qquad\left.-\Pr_{\ket{\phi}\leftarrow\textsf{Haar}(\mathbb{C}^{2^n})}
            \left[\adv\left(\ket{\phi}^{\otimes t(\lambda)}\right)=1\right]\right|
            \leq\negl[\lambda].
        \end{align*} 
    \end{itemize}
{ We focus on two regimes of $\PRS$, based on the state size $n$:}
    \begin{enumerate}
        \item $n= \Theta(\log(\lambda))$, which we call \emph{short pseudorandom states ($\SPRS$s)}.
        \item $n=\Theta (\lambda)$, which we call \emph{long pseudorandom states ($\LPRS$s)}.
    \end{enumerate}  
\end{definition}

{ We also define $\PRG$s, allowing quantum evaluation with negligible correctness error averaged over uniform seeds.}

\begin{definition}[Pseudorandom Generator]
    Let $\lambda\in \mathbb{N}$ be the security parameter and let $n=n(\lambda)$ be polynomial in $\lambda$.{ A QPT algorithm $G$ mapping $\{0,1\}^\lambda$ to $\{0,1\}^n$ is called an $n$-\emph{pseudorandom generator ($\PRG$)} if} 
     \begin{itemize}
     \item \emph{(Expansion)} $n>\lambda$ for all $\lambda\in \mathbb{N}$.
        \item \emph{(Correctness)}{ There is a negligible bound such that, for each $\lambda$, there are strings $\{y_k\in\{0,1\}^n\}_{k\in\{0,1\}^\lambda}$ satisfying} 
        \begin{align}
            \Pr_{k\gets \{0,1\}^\lambda}[G(k)=y_k]\geq 1-\negl[\lambda].
        \end{align}
        \item \emph{(Security)} For any QPT distinguisher $\adv$:
        \begin{align*}
            \left|  \Pr_{{k}\leftarrow \{0,1\}^\lambda} \left[\adv (G(k))=1\right]-\Pr_{y\leftarrow \{0,1\}^n} \left[\adv (y)=1\right]\right| \leq \negl[\lambda].
        \end{align*}
    \end{itemize}
\end{definition}

We now recall the definition of $\botPRG$s from \cite{BBO+24}. First, we define the following useful operator.

\begin{definition}[$\isbot$]
We define the operator 
\begin{align*}
    \isbot(a,b):=\begin{cases}
    \bot        & \text{if } a = \bot \\
        b        & \text{otherwise}.
    \end{cases}
\end{align*}
\end{definition}

\begin{definition}[$\bot$-Pseudorandom Generator]
\label{def:botprg}
    Let $\lambda\in \mathbb{N}$ be the security parameter and let $n=n(\lambda)$ be polynomial in $\lambda$. A QPT algorithm $ G$ mapping $\{0,1\}^\lambda$ to $\{0,1\}^n \cup\{\bot\}$, is a \emph{$(\mu,n)$-$\bot$-pseudodeterministic pseudorandom generator ($\botPRG$)} if:
    \begin{enumerate}
        \item \emph{(Expansion)} $n(\lambda)>\lambda$ for all $\lambda \in \mathbb{N}$.
        \item \emph{(Pseudodeterminism/Correctness)}{ There exists a constant $c>0$ such that $\mu(\lambda)= O(\lambda^{-c})$ and for sufficiently large $\lambda\in \mathbb{N}$ there exists a set $\mathcal{G}_\lambda \subseteq\{0,1\}^\lambda$ satisfying the following conditions. The negligible bounds below are uniform over the indicated keys:}
        \begin{enumerate}
            \item \[\Pr_{k\gets \{0,1\}^\lambda}\left[k\in\mathcal{G}_\lambda \right] \geq 1-\mu(\lambda).\] 
            \item For every $k\in \mathcal{G}_\lambda$ there exists a non-$\bot$ value $y\in\{0,1\}^n$ such that: 
            \begin{align}
                \Pr\left[G(k)=y \right] \geq 1 - \negl[\lambda].
            \end{align} 

            \item For every $k\in \{0,1\}^\lambda$, there exists a non-$\bot$ value $y\in\{0,1\}^n$ such that: 
            \begin{align}
                \Pr\left[G(k)\in \{y,\bot\} \right] \geq 1 - \negl[\lambda].
            \end{align}  
            \end{enumerate}
        \item \emph{(Security)} For every polynomial $q=q(\lambda)$ and QPT distinguisher $\adv$, there exists a negligible function $\epsilon$ such that: 
\begin{align*}
        &\left| \Pr \left[ \begin{matrix}
            k\gets \{0,1\}^\lambda\\
            y_1\gets G(k)\\
            \vdots \\
            y_q \gets G(k)      
        \end{matrix} : \adv(y_1,...,y_q) = 1 \right]\right.\\
        &\qquad\left.- \Pr \left[ \begin{matrix}
            k\gets \{0,1\}^\lambda \\
            y\gets \{0,1\}^{n} \\
            y_1\gets \isbot(G(k),y)\\
            \vdots \\
            y_q \gets \isbot(G(k),y)      
        \end{matrix}: \adv(y_1,\ldots,y_q) = 1    
        \right] \right| \leq \epsilon(\lambda)
    \end{align*}
    \end{enumerate}
\end{definition}

{$\botPRG$s were constructed from $\SPRS$s in \mbox{\cite{BBO+24}}.}

\begin{lemma}[Corollary 1 \cite{BBO+24}]
\label{lem:bot-prg-from-prs}
     {If there exists $(c\log\lambda)$-$\SPRS$ for some constant $c>12$, then there exists a}    $(O(\lambda^{-c/12 +1}),\lambda^{c/12})\text{-}\botPRG$.
\end{lemma}

\subsection{Common Haar Function-like State Oracle}

We first recall the definition of a swap unitary \cite{CCS24}. 

\begin{definition}
\label{def:SWAP}
{ For an $n$-qubit unit vector $\ket{\phi}$ orthogonal to $\ket{0^n}$, the swap (or reflection) unitary is defined by}
\[
S_{\ket{\phi}} = \ket{0^n}\!\bra{\phi} + \ket{\phi}\!\bra{0^n} + I_{\perp} 
\]
{where we assume w.l.o.g.\ that $\ket{\phi}$ is orthogonal to $\ket{0^n}$, as we can always add $\ket{1}$ to make it orthogonal. $I_{\perp}$ is the identity on the subspace orthogonal to $\mathrm{span}\{\ket{0^n}, \ket{\phi}\}$.}\end{definition}

We now define the (unitarized) \emph{Common Haar Function-Like State (\textsf{CHFS}) oracle} as in \cite{BHM+25}. 

\begin{definition}[CHFS oracle]
\label{def:CHFS}
{ Let $\ell=\ell(\lambda)$ be a positive integer-valued, efficiently computable, polynomially bounded function, also defined at $\lambda=0$. We denote by $\AO$ the family of unitary oracles obtained from all choices of the vectors below. Writing $\mathcal{O}\leftarrow\AO$ means sampling those vectors independently according to the Haar distribution:}
\begin{itemize}
    \item \textbf{Randomness:}{ Independently sample $\ket{\phi_x}\leftarrow\textsf{Haar}(\mathbb{C}^{2^{\ell(|x|)}})$ for each $x\in\{0,1\}^*$ and define} 
    \[
    \Phi = 
    \left\{ \ket{\phi_x}\ket{1} \right\}_{x \in \{0,1\}^*}.
    \]
    
    \item \textbf{Setup:} A family of oracles $\mathcal{O}^{\Phi} \coloneqq (S_x^{\Phi})_{x \in \{0,1\}^*} \leftarrow \AO$ is chosen by randomly sampling $\Phi$, where $S_x^{\Phi} := S_{\ket{\phi_x}\ket{1}}$ denotes the swap unitary as defined in \cref{def:SWAP}.
    
    \item \textbf{Query:} The oracle takes as a query a quantum state $\rho_{XYZ}$ such that $\lvert Y \rvert = \ell(\lvert X \rvert) + 1$ and applies the unitary
    {
    \begin{align*}
    \mathcal{O}^{\Phi}
    &:= \sum_{x \in \{0,1\}^{\lvert X \rvert}} \ket{x}\!\bra{x}_X \otimes S_x^{\Phi}\\
    &= \sum_{x \in \{0,1\}^{\lvert X \rvert}} \ket{x}\!\bra{x}_X \otimes S_{\ket{\phi_x}\ket{1}},
    \end{align*}
    }
{ on registers $XY$ of $\rho_{XYZ}$, leaving $Z$ unchanged, where $S_x^{\Phi}$ acts on $Y$. Here $|X|$ and $|Y|$ denote the numbers of qubits in these registers.}
\end{itemize}
\end{definition}
\begin{remark}
    For simplicity, we often discard the superscript $\Phi$ in $\mathcal{O}^\Phi$. Furthermore, for a classical string $x\in \{0,1\}^*$, we write $\mathcal{O}(x)$ to denote the query $\mathcal{O}(\ket{x}\ket{0^{\ell(\lvert x\rvert)+1}})$.

{ This query returns $\ket{x}\ket{\phi_x}\ket{1}$; discarding the label and flag leaves the state $\ket{\phi_x}$.}
\end{remark} 

\subsection{Black-Box Separation}

Black-box separating primitives using oracles was first considered in \cite{IR89} and later formalized in the quantum setting in \cite{CM24}. {We briefly define the relevant notions for this work.}

\begin{definition}
A \emph{primitive} $P$ is a pair $P = (\mathcal{F}_P , \mathcal{R}_P )$ \footnote{We can think of $\mathcal{F}_P$ to mean the ``correctness'' conditions of $P$ and $\mathcal{R}_P$ to mean the ``security'' conditions of $P$.} where $\mathcal{F}_P$ is a set of quantum channels, and $\mathcal{R}_P$ is a relation over pairs $(G, \adv)$ of quantum channels, where $G \in \mathcal{F}_P$. 

A quantum channel $G$ is an \emph{implementation} of $P$ if $G \in \mathcal{F}_P$. If $G$ is additionally a QPT channel, then
we say that $G$ is an \emph{efficient implementation} of $P$.
A quantum channel $\adv$ \emph{$P$-breaks} $G \in \mathcal{F}_P$ if $(G, \adv) \in \mathcal{R}_P$. We say
that $G$ is a \emph{secure implementation} of $P$ if $G$ is an implementation of $P$ such that no QPT channel $P$-breaks
it. The primitive $P$ \emph{exists} if there exists an efficient and secure implementation of $P$.

{ A unitary $U$ implements $P$ if it implements some channel in $\mathcal{F}_P$. We say that $\adv$ $P$-breaks $U$ when it breaks that channel.}
\end{definition}

We now formalize the notion of constructions relative to an oracle.

\begin{definition}
    We say that a primitive $P$ exists relative to an oracle $\mathcal{O}$ if:
    \begin{itemize}
        \item{ There exists a QPT oracle algorithm $G^{(\cdot)}$ such that $G^\mathcal{O}\in \mathcal{F}_P$.}
        \item{ The security of $G^\mathcal{O}$ holds against all QPT adversaries with access to $\mathcal{O}$, i.e., for every QPT oracle algorithm $\adv^{(\cdot)}$, $(G^\mathcal{O},\adv^\mathcal{O})\notin \mathcal{R}_P$.}
    \end{itemize}
\end{definition}

We are now ready to define the notion of fully black-box construction.

{ The correctness requirement below applies to every implementation of $P$, including inefficient and insecure ones. Its negligible correctness bound may depend on the fixed implementation $U$; one bound for all $U$ is not required.}   

\begin{definition}
\label{def:BB with  access to inverse}
A QPT algorithm $G^{(\cdot)}$ is a fully black-box construction (\textsf{BB}) of $Q$ from $P$ \textbf{with inverse access} if the following two conditions hold:
\begin{enumerate}
    \item For every unitary implementation $U$ of $P$, $G^{U,U^{-1}}\in \mathcal{F}_Q$.
\item There is a QPT algorithm $S^{(\cdot)}$ such that, for every
unitary implementation $U$ of $P$, every adversary $\adv$ that $Q$-breaks $G^{U,U^{-1}}$, and every unitary implementation
$\tilde{\adv}$ of $\adv$, it holds that $S^{\tilde{\adv},\tilde{\adv}^{-1}}$ $P$-breaks $U$.
\end{enumerate} 
\end{definition}

The following result from \cite{CM24} shows the relation between \textsf{BB} constructions and oracle separations. 

\begin{theorem}[Theorem 4.2 in \cite{CM24}]
\label{thm:separation relative to unitary}
    Assume there exists a \textsf{BB} construction of a primitive $Q$ from a primitive $P$ with inverse access. Then, for any unitary $\mathcal{O}$, if $P$ exists relative to $(\mathcal{O},\mathcal{O}^{-1})$, then $Q$ exists relative to $(\mathcal{O},\mathcal{O}^{-1})$.
\end{theorem}

\endgroup

%% file: first_separation_corrected_redline.tex
\begingroup
\emergencystretch=3em
\section{Separation}

In this section, we present our separation of $\PRG$ from $\PRS$.

\begin{theorem}
\label{thm:main}
{ Let $\ell=\ell(\lambda)\in O(\lambda)$ be a positive integer-valued, efficiently computable function of the security parameter $\lambda\in \mathbb{N}$ and let $n=n(\lambda)$ be an integer-valued polynomial satisfying $n>\lambda$. There does not exist a \textsf{BB} construction of an $n$-$\PRG$ from an $\ell$-$\PRS$ with inverse access.} 
\end{theorem}

\begin{proof}
\begingroup
{ Our approach is to apply any proposed \textsf{BB} construction to a $\PRS$ based on a unitary oracle. We show that the construction's correctness and security requirements cannot both hold.}
\endgroup

{ Our proof is relative to two unitary oracles: a fixed (unitarized) oracle $\mathcal{C}$ for a \textsf{PSPACE}-complete language and an $\ell$-\textsf{CHFS} oracle $\mathcal{O}$ sampled from a set of oracles $\AO$. The oracle $\mathcal{C}$ is independent of $\mathcal{O}$.} Given that both oracles are self-inverse, it is sufficient to give access to $\mathcal{T}\coloneqq(\mathcal{C},\mathcal{O})$.

{ Assume for contradiction that there exists a \textsf{BB} construction of an $n$-$\PRG$ from an $\ell$-$\PRS$ with inverse access.} First, we state the following result which follows directly from Theorem 4.4 in \cite{BHM+25}.

\begin{Claim}
\label{claim:exist}
{ There exists an $\ell$-$\PRS$ relative to $\mathcal{T}$. Its unitary implementation and inverse only use oracle access to $\mathcal{O}$. It satisfies security with probability 1 over the distribution of $\mathcal{O}$ and correctness for all $\mathcal{O}\in \AO$.} 
\end{Claim}

{
\begin{proof}
On input $k\in\{0,1\}^{\lambda}$, query $\mathcal{O}$ on
$\ket{k}\ket{0^{\ell(\lambda)+1}}$ to obtain
$\ket{k}\ket{\phi_k}\ket{1}$, and output $\ket{\phi_k}$.
The unitary implementation and its inverse each use one query to
$\mathcal{O}$. Correctness holds for every choice of the oracle states.
Security follows from Theorem 4.4 in \cite{BHM+25}, with key length
$\lambda$ and no function input, including access to $\mathcal{C}$.
\qed
\end{proof}
}

{ Applying the assumed \textsf{BB} construction to the $\PRS$ in Claim \ref{claim:exist} gives a QPT generator $\overline{G}:\{0,1\}^\lambda\rightarrow \{0,1\}^{n}$ that only queries $\mathcal{O}$. By the \textsf{BB} correctness requirement, $\overline{G}^{\mathcal{O}}$ satisfies correctness for every $\mathcal{O}\in\AO$, with a negligible bound that may depend on $\mathcal{O}$. By the security reduction, it is secure relative to $\mathcal{T}$ whenever the underlying $\PRS$ is secure, and hence with probability 1 over the distribution of $\mathcal{O}$.} \begingroup
Assume $\overline{G}$ queries the oracle at most $T=T(\lambda)\geq1$ times for some polynomial $T$, and define
\[
 m=m(\lambda)\coloneqq\lceil4\pi T(\lambda)\lambda\rceil+1,
 \qquad
 \delta(\lambda)\coloneqq\frac{1}{100\lambda m(\lambda)}.
\]
We first show that fixing finitely many oracle entries gives a correctness bound that holds for every completion of those entries. For simplicity, a \emph{prefix} fixes all states defining the oracle at lengths $d<a$ for some constant $a$, and a \emph{completion} is any oracle in the family that agrees with these fixed states.
\endgroup

\begin{Claim}
\label{claim:fixed-prefix}
There exist constants $a,\lambda_0$ and fixed states
$\{\ket{\phi_x^*}\}_{|x|<a}$,
{ each obtained by normalizing a nonzero vector whose
coordinates have rational real and imaginary parts,}
such that every $\mathcal{O}\in\AO$ with these entries satisfies,
for every $\lambda\geq\lambda_0$,
\begin{align}
\label{eq:prefix-correctness}
 e_{\mathcal{O}}(\lambda)
 \coloneqq
 \mathbb{E}_{k\gets\{0,1\}^{\lambda}}
 \left[1-\max_y\Pr[\overline{G}^{\mathcal{O}}(k)=y]\right]
 \leq\delta(\lambda).
\end{align}
\end{Claim}

\begin{proof}
\begingroup
For every fixed $\mathcal{O}\in\AO$, correctness implies that
$e_{\mathcal{O}}$ is negligible. For each
$N\in\mathbb{N}$, let
\[
 F_N\coloneqq
 \{\mathcal{O}\in\AO:
 e_{\mathcal{O}}(\lambda)\leq\delta(\lambda)
 \text{ for every }\lambda\geq N\}.
\]
Let $\AO'\subseteq\AO$ consist of the oracles whose states, at every
label, are obtained by normalizing nonzero vectors with rational
real and imaginary coordinates.

First, we show that there exist $a,N$ and a prefix of this form
such that every completion of the prefix in $\AO'$ belongs to
$F_N$. Suppose otherwise.
Starting with the empty prefix and $a_0=0$, we choose strictly
increasing parameters $\lambda_1,\lambda_2,\ldots$ and fix longer
prefixes as follows. At step $j$, choose a completion
$\mathcal{O}^j\in\AO'$ of the already fixed prefix and a parameter
$\lambda_j\geq j$, larger than all previous parameters, such that
\[
 e_{\mathcal{O}^j}(\lambda_j)>\delta(\lambda_j).
\]
Such a choice exists by our supposition. Choose
$a_j>\max\{a_{j-1},j\}$ larger than every query length that can be submitted by $\overline G$ 
when run on seed of length $\lambda_j$.
This is possible because $\overline{G}$ runs in polynomial time.
Fix all entries with $|x|<a_j$ to their values in $\mathcal{O}^j$.
No previously fixed entry changes. 

Since $a_j>j$, the successive prefixes define one oracle
$\mathcal{O}^*\in\AO'\subseteq\AO$, and for every $j$,
\[
 e_{\mathcal{O}^*}(\lambda_j)
 =e_{\mathcal{O}^j}(\lambda_j)
 >\delta(\lambda_j).
\]
This contradicts correctness for $\mathcal{O}^*$, since
$e_{\mathcal{O}^*}$ is negligible and $\delta$ is inverse polynomial.
Thus there exist $a,N$ and a fixed prefix of the required form
whose every completion in $\AO'$ belongs to $F_N$.

We now allow arbitrary states outside this prefix. Fix any
completion $\mathcal{O}\in\AO$ and any $\lambda\geq N$.
For each integer $r\geq1$, approximate the state vector at every
label that $\overline{G}$ can query at this parameter to Euclidean
error at most $1/r$, using normalized vectors with rational real
and imaginary coordinates. Such approximations exist by approximating
each coordinate by rational numbers and then normalizing.
{ Keep the
fixed prefix unchanged, and set all remaining unfixed entries to
$\ket{0^{\ell(|x|)}}$.} This gives a completion
$\mathcal{O}^{(r)}\in\AO'$ with
$e_{\mathcal{O}^{(r)}}(\lambda)\leq\delta(\lambda)$.

The swap definition gives
\[
 \left\|S_{\ket{\phi}\ket{1}}-S_{\ket{\psi}\ket{1}}\right\|_{\mathrm{op}}
 \leq4\left\|\ket{\phi}-\ket{\psi}\right\|_2.
\]
Since the oracle acts separately on each label, the same bound
applies to superposition queries, including queries entangled with
other registers. Replacing the at most $T(\lambda)$ queries one
at a time changes the probability of any output event by at most
$4T(\lambda)/r$. This also covers inverse queries, since the swap
oracles are self-inverse. Therefore,
\[
 e_{\mathcal{O}}(\lambda)
 \leq e_{\mathcal{O}^{(r)}}(\lambda)+\frac{4T(\lambda)}{r}
 \leq\delta(\lambda)+\frac{4T(\lambda)}{r}.
\]
Letting $r$ tend to infinity gives
$e_{\mathcal{O}}(\lambda)\leq\delta(\lambda)$.
This holds for every completion $\mathcal{O}\in\AO$ of the prefix
and every $\lambda\geq N$. Taking $\lambda_0=N$ proves the claim.
\qed
\endgroup
\end{proof}

From now on, $\AO$ denotes the family of completions of this fixed prefix, and $\mathcal{O}\gets\AO$ means that the remaining entries are sampled independently from the Haar distribution.{ The prefix may depend on $\overline{G}$. }

{
The rational descriptions of the fixed states can be hardwired
into an algorithm, since $a$ is constant and states have rational amplitudes. To simulate their swaps,
we use the observation that a swap is a reflection
\cite[Section~5]{GZ25}. Namely, for each $|x|<a$, let
\[
 \ket{v_x}\coloneqq
 \frac{\ket{0^{\ell(|x|)+1}}-\ket{\phi_x^*}\ket{1}}{\sqrt{2}}.
\]
Then
\[
 S_{\ket{\phi_x^*}\ket{1}}
 =I-2\ket{v_x}\!\bra{v_x}.
\]
The known coordinates determine a unitary $V_x$ satisfying
$V_x\ket{0^{\ell(|x|)+1}}=\ket{v_x}$.
Since its dimension is constant, the constructive
Solovay--Kitaev algorithm
\cite[Theorem~1 and Sections~3 and~5]{DN05}
allows us to compute a circuit $\widetilde{V}_x$ approximating
$V_x$ to operator-norm error $\eta/2$, in time polynomial in
$\log(1/\eta)$.

Claim~\ref{claim:exist} still holds under this distribution. Given a QPT adversary for the restricted family, simulate each query on a label $|x|<a$ using the fixed prefix to error $2^{-\lambda}$, and use the original oracle on all other labels. Each query has a fixed input length, so this replaces the entire query when that length is less than $a$ and leaves it unchanged otherwise. The simulation is coherent, and inverse queries are handled by reversing the simulation circuit. Access to $\mathcal{C}$ is unchanged. A polynomial number of queries gives negligible total error. }

Define the oracle $I$ to be the element of $\AO$ whose state is $\ket{\phi_x^*}$ for $|x|<a$ and $\ket{0^{\ell(\lvert x\rvert)}}$ otherwise.{ Queries to $I$ can be approximated to negligible error $\eta$ in polynomial time without any oracle access.}

\begingroup
\begin{lemma}
\label{claim:main}
For every $\mathcal{O}\in\AO$ and sufficiently large $\lambda$, all but a $\frac{1}{10\lambda}$ fraction of inputs $k\in\{0,1\}^{\lambda}$ have a string $y_k$ such that
\begin{align}
\label{eq:23}
 \Pr[\overline{G}^{\mathcal{O}}(k)=y_k]&\geq\frac{9}{10},\\
\label{eq:24}
 \Pr[\overline{G}^{I}(k)=y_k]&\geq\frac{9}{10}.
\end{align}
\end{lemma}
\begin{proof}
Fix $\mathcal{O}\in\AO$ and $\lambda\geq\max\{\lambda_0,2\}$. We define a sequence of $m$ oracles $\mathcal{O}_1,\ldots,\mathcal{O}_m$ starting with $\mathcal{O}_1=\mathcal{O}$ and ending with $\mathcal{O}_m=I$. For every label $x$, gradually move its state vector along a shortest path on the unit sphere from its value in $\mathcal{O}$ to its value in $I$. We use the full vectors, including their phases, and keep the entries with $|x|<a$ unchanged. Each path has length at most $\pi$. Splitting it into $m-1$ equal steps gives consecutive vectors at Euclidean distance at most $\pi/(m-1)$.

For any two unit vectors $\ket{\phi},\ket{\psi}$, the corresponding swaps satisfy
{
\[
 \left\|S_{\ket{\phi}\ket{1}}-S_{\ket{\psi}\ket{1}}\right\|_{\mathrm{op}}
 \leq4\left\|\ket{\phi}-\ket{\psi}\right\|_2.
\]
}
{ This follows directly from the swap definition in \cref{def:SWAP}.} Since the oracle acts separately on each label, it follows that
\[
 \|\mathcal{O}_i-\mathcal{O}_{i+1}\|_{\mathrm{op}}
 \leq\frac{4\pi}{m-1}.
\]
Replacing the at most $T$ queries one at a time therefore gives, for every $i\in[m-1]$ and every input $k$,
\begin{align}
\label{eq:trace-dis}
 \textsf{Tr}\left(\overline{G}^{\mathcal{O}_i}(k),
                  \overline{G}^{\mathcal{O}_{i+1}}(k)\right)
 \leq\frac{4\pi T}{m-1}\leq\frac{1}{\lambda}.
\end{align}
The same bound holds for inverse queries, since the swap oracles are self-inverse.

Every $\mathcal{O}_i$ belongs to $\AO$, so \cref{eq:prefix-correctness} applies to it. By Markov's inequality, at most a $10\delta(\lambda)$ fraction of inputs have no output occurring with probability at least $9/10$ under $\overline{G}^{\mathcal{O}_i}$. Taking a union bound over the $m$ oracles, all but a
\[
 10m\delta(\lambda)=\frac{1}{10\lambda}
\]
fraction of inputs have such an output at every intermediate oracle.

Fix any remaining input $k$. Its most likely output cannot change between consecutive oracles: two distinct outputs, each occurring with probability at least $9/10$ in its respective distribution, would give trace distance at least $4/5$, contradicting \cref{eq:trace-dis}. Hence the same string $y_k$ satisfies \cref{eq:23,eq:24}.
\qed
\end{proof}
\endgroup

\begingroup
We now give a search attack against $\overline{G}$ using the \textsf{PSPACE} oracle $\mathcal{C}$. {Since $I$ can be efficiently approximated, $\overline{G}^{I}$ can be simulated without any oracle access. A \textsf{PSPACE} computation can therefore estimate $\Pr[\overline{G}^{I}(k)=y]$ for every seed $k$.}

{On input $y$, let $\adv^{\mathcal{C}}$ output $1$ if some $k\in\{0,1\}^{\lambda}$ has an estimated probability at least $2/3$, and output $0$ otherwise.} The search over all seeds uses polynomial space, so $\adv$ can perform it with a query to $\mathcal{C}$.

For each seed, at most one string can pass this test. Thus at most $2^\lambda$ strings are accepted, and
\begin{align}
\label{eq:uniform-accept}
 \Pr_{y\gets\{0,1\}^{n}}[\adv^{\mathcal{C}}(y)=1]
 \leq2^{\lambda-n}\leq\frac12.
\end{align}

On the other hand, fix any $\mathcal{O}\in\AO$. By \cref{claim:main}, all but a $\frac{1}{10\lambda}$ fraction of seeds have the same most likely output $y_k$ under $\overline{G}^{\mathcal{O}}$ and $\overline{G}^{I}$, with probability at least $9/10$ in both cases. This string passes the test. By \cref{eq:prefix-correctness}, the average probability that $\overline{G}^{\mathcal{O}}$ fails to output its most likely string is at most $\delta(\lambda)$. Therefore,
\begin{align}
\label{eq:real-accept}
 \Pr_{k\gets\{0,1\}^{\lambda}}
 [\adv^{\mathcal{C}}(\overline{G}^{\mathcal{O}}(k))=1]
 \geq1-\frac{1}{10\lambda}-\delta(\lambda).
\end{align}
Combining \cref{eq:uniform-accept,eq:real-accept}, the distinguishing advantage is at least
\begin{align}
\label{eq:322}
 \frac12-\frac{1}{10\lambda}-\delta(\lambda),
\end{align}
which is bounded below by a positive constant for sufficiently large $\lambda$. This attack works for every $\mathcal{O}\in\AO$, contradicting the security of $\overline{G}$ whenever the underlying $\PRS$ is secure.
\endgroup

{ Therefore, there does not exist a \textsf{BB} construction of an $n$-$\PRG$ from an $\ell$-$\PRS$ with inverse access.}
\qed
\end{proof}

\endgroup

%% file: second_separation_corrected_redline.tex
\begingroup
\emergencystretch=3em
\section{Separating PRG from $\bot$-PRG}

{
To demonstrate the separation between $\PRG$s and $\botPRG$s, we will use two independent unitary oracles: a fixed (unitarized) oracle $\mathcal{C}$ for a \textsf{PSPACE}-complete language and a $\bot$-pseudodeterministic random oracle $\mathcal{O}$ described in Construction \ref{con:oracles 1}, similar to one used in \cite{BBO+24}.
}


\begin{construct}
\label{con:oracles 1}
Let $\lambda\in\mathbb{N}$ be the security parameter, let $c>0$
be a constant, and set $\mu(\lambda)=\lambda^{-c}$.
{
Let $\ell$ be an integer-valued polynomial with $\ell(\lambda)>\lambda$.
}

Sample $O_\lambda\leftarrow\Pi_{\lambda,\ell}$.
Independently, sample a function
$Q_\lambda:\{0,1\}^\lambda\to[0,1]$ by choosing each value
uniformly from
\[
    \left\{\frac{j}{2^\lambda}:
    j=0,\ldots,2^\lambda-1\right\}.
\]
Sample a uniformly random set
$\mathcal{P}_\lambda\subseteq\{0,1\}^\lambda$
of size $\lfloor\mu(\lambda)2^\lambda\rfloor$,
and let $P_\lambda$ be its indicator function.

For each $x\in\{0,1\}^\lambda$, define
\[
    p_x\coloneqq
    \sin^2\left(\frac{\pi}{2}Q_\lambda(x)\right),
    \qquad
    \ket{\phi_x}\coloneqq
    \begin{cases}
        \sqrt{p_x}\ket{\bot}
        +\sqrt{1-p_x}\ket{O_\lambda(x)},
        & P_\lambda(x)=1,\\
        \ket{O_\lambda(x)},
        & P_\lambda(x)=0.
    \end{cases}
\]
Define the unitary oracle
\[
    \mathcal{O}_\lambda
    \coloneqq \mathcal{O}[P_\lambda,Q_\lambda,O_\lambda]
    \coloneqq
    \sum_{x\in\{0,1\}^\lambda}
        \ket{x}\!\bra{x}\otimes S_{\ket{\phi_x}},
\]
{
where $\ket{0}$, $\ket{\bot}$, and the encoded output strings $\ket{y}$ are distinct computational-basis states, so that the initialization state $\ket{0}$ is orthogonal to every possible output, including the encoded string $\ket{0^{\ell(\lambda)}}$,
}
\[
    S_{\ket{\phi_x}}
    \coloneqq
    \ket{0}\!\bra{\phi_x}
    +\ket{\phi_x}\!\bra{0}
    +I_\perp,
\]
and $I_\perp$ is the identity on the subspace orthogonal to
$\operatorname{span}\{\ket{0},\ket{\phi_x}\}$.
Thus,
\[
    \mathcal{O}_\lambda
        \bigl(\ket{x}\ket{0}\bigr)
    =\ket{x}\ket{\phi_x},
    \qquad
    \mathcal{O}_\lambda^{-1}=\mathcal{O}_\lambda.
\]
\end{construct}

We define the ``good'' set $\mathcal{G}_\lambda^\mathcal{O}$ for $\mathcal{O}_\lambda$ as follows: 
\begin{align*}
    \mathcal{G}_\lambda^\mathcal{O}\coloneqq \{x\in \{0,1\}^\lambda: P_\lambda(x)=0\}.
\end{align*}

We let $\mathcal{O}_{\mu,\ell}$ denote the family of oracles
$(\mathcal{O}_\lambda)_{\lambda\in\mathbb{N}}$ defined above,
allowing arbitrary functions $O_\lambda$, arbitrary functions
$Q_\lambda:\{0,1\}^\lambda\to[0,1]$, and arbitrary indicator
functions $P_\lambda$ satisfying
\[
    \bigl|\{x:P_\lambda(x)=1\}\bigr|
    \leq \lfloor\mu(\lambda)2^\lambda\rfloor.
\]
Sampling $\mathcal{O}\leftarrow\mathcal{O}_{\mu,\ell}$
means using the distribution in Construction~\ref{con:oracles 1},
independently at each input length.

When we write $\mathcal{O}_\lambda(x)$ as a classical output,
we mean the outcome obtained by applying the oracle to
$\ket{x}\ket{0}$ and measuring the output register.
The measurement is performed by the calling algorithm.


The following lemma follows directly from the definition of $\mathcal{O}$.  
\begin{lemma}
\label{lem:O}
    $\mathcal{O}_\lambda$ has the following properties:
\begin{itemize}
            \item $\Pr_{x\leftarrow \{0,1\}^\lambda}\left[ x\in \mathcal{G}_\lambda^\mathcal{O}\right] \geq 1-\mu.$

  \item For every $x\in \mathcal{G}_\lambda^\mathcal{O}$, there exists a non-$\bot$ value $y\in\{0,1\}^{\ell}$ such that: 
  \begin{align*}
      \Pr\left[\mathcal{O}_\lambda(x)=y \right] =1.
  \end{align*}
            \item For every $x\notin \mathcal{G}_\lambda^\mathcal{O}$, there exists a probability $p_x\in [0,1]$ and non-$\bot$ value $y\in\{0,1\}^\ell$ such that: 
            \begin{enumerate}
                \item  $\Pr\left[ \mathcal{O}_\lambda(x)=y \right] =1-p_x.$ 
            \item $\Pr\left[  \mathcal{O}_\lambda(x)=\bot \right] =p_x$. 
            \end{enumerate}            
\end{itemize}
\end{lemma}


\begin{theorem}
{
Let $\lambda\in \mathbb{N}$ be the security parameter. For any integer-valued polynomials $w(\lambda),\ell(\lambda)>\lambda$ and pseudodeterminism error $\mu(\lambda)=\lambda^{-c}$ for $c>0$, there does not exist a \textsf{BB} construction of a $w$-$\PRG$ from a $(3\mu,\ell)$-$\botPRG$ with inverse access.
} 
\end{theorem}

\begin{proof}
\begingroup
{
Our approach is to apply any proposed \textsf{BB} construction to a $\botPRG$ based on the self-inverse unitary oracles $\mathcal{T}\coloneqq(\mathcal{C},\mathcal{O})$. We show that the construction's correctness and security requirements cannot both hold.
}
\endgroup

{
Assume for contradiction that there exists a \textsf{BB} construction of a $w$-$\PRG$ $\overline{G}^F$ from a $(3\mu,\ell)$-$\botPRG$ $F$, with inverse access.
} First, we show that there exists a $(\mu,\ell)$-$\botPRG$ relative to the oracles $(\mathcal{O},\mathcal{C})$.

\begin{Claim}
\label{lem:determinism}
{
For every fixed admissible pair of sequences $P,Q$ chosen independently of $O$, with probability $1$ over $O$, the generator that evaluates $\mathcal{O}_\lambda=\mathcal{O}[P_\lambda,Q_\lambda,O_\lambda]$ and measures its output is a $(\mu,\ell)$-$\botPRG$ relative to $(\mathcal{O},\mathcal{C})$. Furthermore, correctness is satisfied for every $\mathcal{O}\in\mathcal{O}_{\mu,\ell}$.
}
\end{Claim}


\begin{proof}
By \cref{lem:O}, $\mathcal{O}$ satisfies the correctness/pseudodeterminism condition of a $(\mu,\ell)$-$\botPRG$. 

For security, we need to show that for any $P,Q$ and with probability 1 over the distribution of $O$: for every non-uniform QPT distinguisher $\adv$ and polynomial $q=q(\lambda)$: 
\begin{align*}
        \left| \Pr \left[ \begin{matrix}
            k\gets \{0,1\}^\lambda\\
            y_1\gets \mathcal{O}_\lambda(k)\\
            \vdots \\
            y_q \gets \mathcal{O}_\lambda(k)      
        \end{matrix} : \adv^{\mathcal{O},\mathcal{C}}(y_1,...,y_q) = 1 \right] -
        \Pr \left[ \begin{matrix}
            k\gets \{0,1\}^\lambda \\
            y\gets \{0,1\}^{\ell} \\
            y_1\gets \isbot(\mathcal{O}_\lambda(k),y)\\
            \vdots \\
            y_q \gets \isbot(\mathcal{O}_\lambda(k),y)      
        \end{matrix}: \adv^{\mathcal{O},\mathcal{C}}(y_1,\ldots,y_q) = 1    
        \right] \right| \leq \negl[\lambda]
    \end{align*}

Let $Z_\lambda$ be the function that outputs $0^\ell$ on any input and let $\mathcal{Z}_\lambda\coloneqq \mathcal{O}[P_\lambda,Q_\lambda,Z_\lambda]$. Note that \begin{itemize}
    \item $\mathcal{Z}_\lambda$ is independent of $O_\lambda$,
    \item $\mathcal{O}_\lambda(k)=\isbot(\mathcal{O}_\lambda(k),O_\lambda(k))=\isbot(\mathcal{Z}_\lambda(k),O_\lambda(k))$,
    \item $\isbot(\mathcal{O}_\lambda(k),y) =\isbot(\mathcal{Z}_\lambda(k),y) $.
\end{itemize}

Therefore, $\adv^{\mathcal{O},\mathcal{C}}$ needs to distinguish between evaluations of $\isbot(\mathcal{Z}_\lambda(k),y)$ and $\isbot(\mathcal{Z}_\lambda(k),O_\lambda(k))$. 

Lemma 2.2 from \cite{SXY18} states that a random oracle acts as a $\PRG$ i.e.:
\begin{align*}
   \underset{{O\leftarrow \Pi_{\lambda,\ell}}}{\mathbb{E}} \left[ \left| \Pr_{k\leftarrow \{0,1\}^\lambda}\left[\adv^{O}(O(k))=1\right]-    \Pr_{y\leftarrow \{0,1\}^\ell}\left[\adv^{O}(y)=1\right]\right| \right]\leq \frac{1}{2^{\lambda/4}}.
\end{align*}

Note that this result even holds against unbounded-time adversaries as long as the number of queries to the oracle is polynomial. Hence, this result also holds against adversaries with access to a \textsf{PSPACE}-oracle:
\begin{align*}
   \underset{{O\leftarrow \Pi_{\lambda,\ell}}}{\mathbb{E}} \left[ \left| \Pr_{k\leftarrow \{0,1\}^\lambda}\left[\adv^{O,\mathcal{C}}(O(k))=1\right]-    \Pr_{y\leftarrow \{0,1\}^\ell}\left[\adv^{O,\mathcal{C}}(y)=1\right]\right| \right]\leq \frac{1}{2^{\lambda/4}}.
\end{align*}

Next, notice that for any functions $P,Q$, distinguishing between evaluations of $\isbot(\mathcal{Z}_\lambda(k),y)$ and $\isbot(\mathcal{Z}_\lambda(k),O_\lambda(k))$ is just as hard as distinguishing the two scenarios in the equation above, given that $\mathcal{Z}_\lambda$ is independent of $O_\lambda$. Therefore, 
\begin{align*}
      \underset{{{O\leftarrow \Pi_{\lambda,\ell}}}}{\mathbb{E}} \left[ \left| \Pr_{(y_1,...,y_q)\leftarrow D_{\mathcal{Z},O}^0} \left[ \adv^{\mathcal{O},\mathcal{C}}(y_1,...,y_q) = 1 \right] - \Pr_{(y_1,...,y_q)\leftarrow D_{\mathcal{Z}}^1} \left[  \adv^{\mathcal{O},\mathcal{C}}(y_1,\ldots,y_q) = 1    
        \right] \right|\right] \leq 2^{-\lambda/4}
    \end{align*}
where, 
\begin{align*}
D_{\mathcal{Z},O}^0\coloneqq \left[
\begin{matrix}
            k\gets \{0,1\}^\lambda\\
            y_1\gets \isbot(\mathcal{Z}_\lambda(k),O_\lambda(k))\\
            \vdots \\
            y_q \gets \isbot(\mathcal{Z}_\lambda(k),O_\lambda(k))     
        \end{matrix}    \right]  \  D_{\mathcal{Z}}^1\coloneqq \left[
\begin{matrix}
            k\gets \{0,1\}^\lambda \\
            y\gets \{0,1\}^{\ell} \\
            y_1\gets \isbot(\mathcal{Z}_\lambda(k),y) \\
            \vdots \\
            y_q \gets \isbot(\mathcal{Z}_\lambda(k),y)       
        \end{matrix}\right]
\end{align*}

By Markov inequality, we get that
\begin{align*}
      \Pr_{{O\leftarrow \Pi_{\lambda,\ell}}} \left[ \left|  \Pr_{(y_1,...,y_q \leftarrow D_{\mathcal{Z},O}^0}\right. \right. & \left[ \adv^{\mathcal{O},\mathcal{C}}(y_1,...,y_q) = 1 \right] -\\ 
      &\left. \left. \Pr_{(y_1,...,y_q)\leftarrow D_{\mathcal{Z}}^1} \left[  \adv^{\mathcal{O},\mathcal{C}}(y_1,\ldots,y_q) = 1   
        \right] \right| \geq {2^{-\lambda/8}}\right] \leq 2^{-\lambda/8}
    \end{align*}
    
By Borel-Cantelli Lemma, since $\sum_\lambda 2^{-\lambda/8}$ converges, with probability 1 over the distribution of $O$, it holds that
\begin{align*}
      \left|  \Pr_{(y_1,...,y_q)\leftarrow D_{\mathcal{Z},O}^0} \left[ \adv^{\mathcal{O},\mathcal{C}}(y_1,...,y_q) = 1 \right] - \Pr_{(y_1,...,y_q)\leftarrow D_{\mathcal{Z}}^1} \left[  \adv^{\mathcal{O},\mathcal{C}}(y_1,\ldots,y_q) = 1    
        \right] \right| \leq {2^{-\lambda/8}},
    \end{align*}

except for finitely many $\lambda\in \mathbb{N}$. There are countably many quantum algorithms $\adv$ making polynomial queries to $(\mathcal{O},\mathcal{C})$, so this bound holds for every such adversary. Therefore, $\mathcal{O}$ is a $\botPRG$ for any $P,Q$ and with probability 1 over the distribution of $O$.  
    \qed
\end{proof}

{
We instantiate $F$ by the generator above. Its unitary implementation and inverse only query $\mathcal{O}$, so the resulting generator $\overline{G}^{\mathcal{O}}$ only queries $\mathcal{O}$. By the assumed \textsf{BB} construction, it satisfies correctness for every $\mathcal{O}\in\mathcal{O}_{3\mu,\ell}$.} Indeed, \cref{lem:O}, with $\mu$ replaced by $3\mu$, gives source correctness throughout this larger family.

Assume $\overline{G}$ makes at most $T$ queries, on inputs of length at most $T$, where $T=T(\lambda)\geq\lambda$ is polynomial. Define
\begin{align*}
 m=m(\lambda)&\coloneqq\lceil2\pi T(\lambda)\lambda\rceil,
 &n=n(\lambda)&\coloneqq\left\lceil\frac{24m(\lambda)}{\mu(T(\lambda))}\right\rceil,\\
 \delta(\lambda)&\coloneqq\frac{1}{100\lambda n(\lambda)}.
\end{align*}
Here $n$ bounds the number of intermediate oracles used below.

\begin{Claim}
\label{claim:fixed-prefix 2}
There exist constants $a,\lambda_0$ and fixed tables $(P_d^*,Q_d^*,O_d^*)_{d<a}$, with rational entries in $Q_d^*$, such that every $\mathcal{O}\in\mathcal{O}_{3\mu,\ell}$ with these functions, for every $\lambda\geq\lambda_0$,
\begin{align}
\label{eq:correctness 2}
 e_{\mathcal{O}}(\lambda)
 \coloneqq
 \mathbb{E}_{k\gets\{0,1\}^{\lambda}}
 \left[1-\max_y\Pr[\overline{G}^{\mathcal{O}}(k)=y]\right]
 \leq\delta(\lambda).
\end{align}
\end{Claim}
\begin{proof}
{
We use the same argument as in Claim~\ref{claim:fixed-prefix}.
First restrict to oracles whose $Q$-entries are rational.
Suppose that no finite prefix with rational $Q$-entries has a
correctness bound $e_{\mathcal{O}}(\lambda)\leq\delta(\lambda)$
for all sufficiently large $\lambda$ and all such completions.

Starting with the empty prefix, choose strictly increasing
parameters $\lambda_j\geq j$ and a completion $\mathcal{O}^j$
of the previously fixed tables such that
\[
 e_{\mathcal{O}^j}(\lambda_j)>\delta(\lambda_j).
\]
Such a choice exists by our supposition. Choose
$a_j>\max\{a_{j-1},j,T(\lambda_j)\}$, with $a_0=0$, and fix
all tables at lengths $d<a_j$ to those of $\mathcal{O}^j$.
Every later completion has the same output distribution at
$\lambda_j$, since all possible query lengths have been fixed.
The successive prefixes therefore define one admissible oracle
$\mathcal{O}^*$ with rational $Q$-entries satisfying
\[
 e_{\mathcal{O}^*}(\lambda_j)>\delta(\lambda_j)
 \qquad\text{for every }j.
\]
This contradicts correctness for $\mathcal{O}^*$, since its error
is negligible and $\delta$ is inverse polynomial. Thus there
exist $a,N$ and a fixed prefix such that every completion with
rational $Q$-entries satisfies the bound for all $\lambda\geq N$.

Now fix any completion $\mathcal{O}\in\mathcal{O}_{3\mu,\ell}$
of this prefix and any $\lambda\geq N$.
For each integer $r\geq1$, replace every $Q$-entry outside the
prefix by a rational value in $[0,1]$ within $1/r$, leaving
$P,O$ and the prefix unchanged. Call the resulting oracle
$\mathcal{O}^{(r)}$. Its correctness error at $\lambda$ is at
most $\delta(\lambda)$.

By Construction~\ref{con:oracles 1}, each prepared state changes
by at most $\pi/(2r)$ in Euclidean distance, and its swap changes
by at most $2\pi/r$ in operator norm. This bound also holds for
superposition queries. Replacing the at most $T(\lambda)$ queries
one at a time, including inverse queries, gives
\[
 e_{\mathcal{O}}(\lambda)
 \leq e_{\mathcal{O}^{(r)}}(\lambda)
       +\frac{2\pi T(\lambda)}{r}
 \leq\delta(\lambda)+\frac{2\pi T(\lambda)}{r}.
\]
Letting $r$ tend to infinity gives the required bound.
Taking $\lambda_0=N$ proves the claim.
}
\qed
\end{proof}

Define for any $d\geq a$,
\[
 b_d\coloneqq\lfloor\mu(d)2^d\rfloor.
\]
We now keep this prefix fixed and sample all remaining entries as in Construction~\ref{con:oracles 1}. {The security conclusion of Claim~\mbox{\ref{lem:determinism}} is unchanged: the finite rational prefix can be hardcoded and simulated to negligible error, while the remaining entries retain their original distribution.} The resulting oracle is a $(3\mu,\ell)$-$\botPRG$, since the fixed prefix satisfies the $3\mu$ budget and all longer lengths satisfy the $\mu$ budget.

Define $\mathcal{Z}$ to retain the fixed prefix and, at every length $d\geq a$, to use $P_d=Q_d=0$ and $O_d(x)=0^{\ell(d)}$. Thus $\mathcal{Z}\in\mathcal{O}_{3\mu,\ell}$ can be approximated efficiently.

\begingroup
\begin{lemma}
\label{claim:main 2}
For every oracle $\mathcal{O}$ with the fixed prefix and the $\mu$ budget at all lengths $d\geq a$, and for sufficiently large $\lambda$, all but a $\frac{1}{10\lambda}$ fraction of inputs $k\in\{0,1\}^{\lambda}$ have a string $y_k$ such that
\begin{align}
\label{eq:232}
 \Pr[\overline{G}^{\mathcal{O}}(k)=y_k]&\geq\frac{9}{10},\\
\label{eq:242}
 \Pr[\overline{G}^{\mathcal{Z}}(k)=y_k]&\geq\frac{9}{10}.
\end{align}
\end{lemma}
\begin{proof}
Fix $\mathcal{O}$ and take $\lambda\geq\max\{a,\lambda_0,2\}$. We first construct a sequence of at most $n$ oracles from $\mathcal{O}$ to $\mathcal{Z}$. We change the classical outputs by first moving them to $\bot$.

For each $q\in[\lceil2/\mu(T)\rceil]$, define
\begin{align}
\label{eq:x}
 S_{T,q}\coloneqq
 \{x\in\{0,1\}^d:\ a\leq d\leq T,\ 
       (q-1)b_d\leq x<qb_d\},
\end{align}
where $x$ is interpreted as an integer. These sets partition all inputs with lengths between $a$ and $T$: at each length $d$, the number of blocks is at most
$\lceil2/\mu(d)\rceil\leq\lceil2/\mu(T)\rceil$.

Starting from the tables $(P,Q,O)$ of $\mathcal{O}$, perform the following steps for each block $S_{T,q}$. Each change applies to every $x$ in that block and leaves all other entries unchanged.
\begin{enumerate}
 \item Repeatedly set $Q(x)\gets\max(Q(x)-1/m,0)$ until $Q(x)=0$ throughout the block.
 \item Set $P(x)\gets1$ throughout the block.
 \item Repeatedly set $Q(x)\gets\min(Q(x)+1/m,1)$ until $Q(x)=1$ throughout the block.
 \item Set $O(x)\gets0^{\ell(|x|)}$ throughout the block.
 \item Repeatedly set $Q(x)\gets\max(Q(x)-1/m,0)$ until $Q(x)=0$ throughout the block.
 \item Set $P(x)\gets0$ throughout the block.
\end{enumerate}
Record the oracle after each change. Finally, replace entries at lengths greater than $T$ by those of $\mathcal{Z}$; $\overline{G}$ never queries these entries.

At each length $d\geq a$, the support of $P$ is contained in its original support together with the current block. Hence it has size at most $2b_d\leq\lfloor3\mu(d)2^d\rfloor$. The prefix is unchanged. Therefore, every intermediate oracle belongs to $\mathcal{O}_{3\mu,\ell}$ and satisfies \cref{eq:correctness 2}.

Each block introduces at most $3m+3$ changes. Including the initial oracle and the final change above length $T$, the number of oracles is at most
\[
 (3m+3)\left\lceil\frac{2}{\mu(T)}\right\rceil+2
 \leq\frac{24m}{\mu(T)}\leq n.
\]
By repeating the final oracle if necessary, write the sequence as $\mathcal{O}_1,\ldots,\mathcal{O}_n$, with $\mathcal{O}_1=\mathcal{O}$ and $\mathcal{O}_n=\mathcal{Z}$.

It remains to bound the distance between consecutive evaluations. When $P(x)=1$, the prepared state is
\[
 \sin\left(\frac{\pi Q(x)}2\right)\ket{\bot}
 +\cos\left(\frac{\pi Q(x)}2\right)\ket{O(x)}.
\]
Changing $Q(x)$ by at most $1/m$ changes this vector by at most $\pi/(2m)$ in Euclidean distance. The corresponding swaps differ by at most $2\pi/m$ in operator norm, using
\[
 \|S_{\ket{\phi}}-S_{\ket{\psi}}\|_{\mathrm{op}}
 \leq4\|\ket{\phi}-\ket{\psi}\|_2.
\]
Changing $P$ at $Q=0$ leaves the oracle unchanged. Changing $O$ at $P=1,Q=1$ also leaves it unchanged, since the prepared state is then $\ket{\bot}$. These bounds hold for superposition queries as well, because the oracle acts separately on each label.

Replacing the at most $T$ queries one at a time therefore gives, for every $i\in[n-1]$ and every input $k$,
\begin{align}
\label{eq:trace-dis 2}
 \textsf{Tr}\left(\overline{G}^{\mathcal{O}_i}(k),
                  \overline{G}^{\mathcal{O}_{i+1}}(k)\right)
 \leq\frac{2\pi T}{m}\leq\frac1\lambda.
\end{align}
The final change above length $T$ has no effect on the computation. The same bounds apply to inverse queries, since the oracles are self-inverse.

By \cref{eq:correctness 2} and Markov's inequality, at most a $10\delta(\lambda)$ fraction of seeds have no output occurring with probability at least $9/10$ at any fixed intermediate oracle. Taking a union bound over the $n$ oracles, all but a
\[
 10n\delta(\lambda)=\frac1{10\lambda}
\]
fraction of seeds have such an output at every intermediate oracle. For each remaining seed, this output cannot change along the sequence: distinct outputs occurring with probability at least $9/10$ in consecutive distributions would give trace distance at least $4/5$, contradicting \cref{eq:trace-dis 2}. Thus the same string satisfies \cref{eq:232,eq:242}.
\qed
\end{proof}
\endgroup

We now give a search attack against $\overline{G}$ using $\mathcal{C}$. {The fixed prefix of $\mathcal{Z}$ has finite rational tables and all its remaining entries are zero. Hence a \textsf{PSPACE} computation can estimate $\Pr[\overline{G}^{\mathcal{Z}}(k)=y]$ and distinguish the output from random. This contradicts the existence of a $\PRG$ in the same way as in the previous separation.}

{
Therefore, there does not exist a \textsf{BB} construction of a $w$-$\PRG$ from a $(3\mu,\ell)$-$\botPRG$ with inverse access.
}
\qed
\end{proof}

\endgroup

%% file: mybib.bib
@misc{BM24,
  author       = {Samuel Bouaziz{-}Ermann and
                  Garazi Muguruza},
  title        = {Quantum Pseudorandomness Cannot Be Shrunk In a Black-Box Way},
  year         = {2024},
  url          = {https://doi.org/10.48550/arXiv.2402.13324},
  eprinttype    = {arXiv},
  eprint       = {2402.13324}
}

@article{KQT24,
  title={Quantum-Computable One-Way Functions without One-Way Functions},
  author={Kretschmer, William and Qian, Luowen and Tal, Avishay},
  journal={arXiv preprint arXiv:2411.02554},
  year={2024}
}

@misc{GZ25,
  author        = {Eli Goldin and Mark Zhandry},
  title         = {Translating Between the Common Haar Random
                   State Model and the Unitary Model},
  year          = {2025},
  eprint        = {2503.11634},
  archivePrefix = {arXiv},
  primaryClass  = {quant-ph},
  url           = {https://arxiv.org/abs/2503.11634}
}

@misc{DN05,
  author        = {Christopher M. Dawson and Michael A. Nielsen},
  title         = {The {Solovay--Kitaev} Algorithm},
  year          = {2005},
  eprint        = {quant-ph/0505030},
  archivePrefix = {arXiv},
  url           = {https://arxiv.org/abs/quant-ph/0505030}
}

@article{CCS24,
  title={The power of a single Haar random state: constructing and separating quantum pseudorandomness},
  author={Chen, Boyang and Coladangelo, Andrea and Sattath, Or},
  journal={arXiv preprint arXiv:2404.03295},
  year={2024}
}

@inproceedings{CM24,
  title={On black-box separations of quantum digital signatures from pseudorandom states},
  author={Coladangelo, Andrea and Mutreja, Saachi},
  booktitle={Theory of Cryptography Conference},
  pages={289--317},
  year={2024},
  organization={Springer}
}

@article{BM84,
author = {Blum, Manuel and Micali, Silvio},
title = {How to Generate Cryptographically Strong Sequences of Pseudorandom Bits},
journal = {SIAM Journal on Computing},
volume = {13},
number = {4},
pages = {850-864},
year = {1984},
doi = {10.1137/0213053},
eprint = { https://doi.org/10.1137/0213053}
}

@inproceedings{K21,
  title={Quantum Pseudorandomness and Classical Complexity},
  author={Kretschmer, William},
  booktitle={16th Conference on the Theory of Quantum Computation, Communication and Cryptography},
  year={2021}
}

@inproceedings{AQY22,
  title={Cryptography from pseudorandom quantum states},
  author={Ananth, Prabhanjan and Qian, Luowen and Yuen, Henry},
  booktitle={Advances in Cryptology--CRYPTO 2022: 42nd Annual International Cryptology Conference, CRYPTO 2022, Santa Barbara, CA, USA, August 15--18, 2022, Proceedings, Part I},
  pages={208--236},
  year={2022},
  organization={Springer}
}

@inproceedings{BS20,
  title={Scalable pseudorandom quantum states},
  author={Brakerski, Zvika and Shmueli, Omri},
  booktitle={Annual International Cryptology Conference},
  pages={417--440},
  year={2020},
  organization={Springer}
}

@inproceedings{SXY18,
  title={Tightly-secure key-encapsulation mechanism in the quantum random oracle model},
  author={Saito, Tsunekazu and Xagawa, Keita and Yamakawa, Takashi},
  booktitle={Advances in Cryptology--EUROCRYPT 2018: 37th Annual International Conference on the Theory and Applications of Cryptographic Techniques, Tel Aviv, Israel, April 29-May 3, 2018 Proceedings, Part III 37},
  pages={520--551},
  year={2018},
  organization={Springer}
}

@inproceedings{BNY25,
  title={Microcrypt assumptions with quantum input sampling and pseudodeterminism: Constructions and separations},
  author={Barhoush, Mohammed and Nishimaki, Ryo and Yamakawa, Takashi},
  booktitle={International Conference on the Theory and Application of Cryptology and Information Security},
  pages={516--548},
  year={2025},
  organization={Springer}
}

@inproceedings{B21,
  title={Secure quantum computation with classical communication},
  author={Bartusek, James},
  booktitle={Theory of Cryptography Conference},
  pages={1--30},
  year={2021},
  organization={Springer}
}

@inproceedings{MY22a,
  title={Quantum commitments and signatures without one-way functions},
  author={Morimae, Tomoyuki and Yamakawa, Takashi},
  booktitle={Advances in Cryptology--CRYPTO 2022: 42nd Annual International Cryptology Conference, CRYPTO 2022, Santa Barbara, CA, USA, August 15--18, 2022, Proceedings, Part I},
  pages={269--295},
  year={2022},
  organization={Springer}
}

@book{NC00,
	address = {New York, NY, USA},
	author = {Nielsen, Michael A. and Chuang, Isaac L.},
	doi = {10.1017/CBO9780511976667},
	isbn = {0-521-63503-9},
	publisher = {Cambridge University Press},
	title = {Quantum computation and quantum information},
	year = {2000}}

@inproceedings{IR89,
  title={Limits on the provable consequences of one-way permutations},
  author={Impagliazzo, Russell and Rudich, Steven},
  booktitle={Proceedings of the twenty-first annual ACM symposium on Theory of computing},
  pages={44--61},
  year={1989}
}

@inproceedings{JLS18,
  title={Pseudorandom quantum states},
  author={Ji, Zhengfeng and Liu, Yi-Kai and Song, Fang},
  booktitle={Advances in Cryptology--CRYPTO 2018: 38th Annual International Cryptology Conference, Santa Barbara, CA, USA, August 19--23, 2018, Proceedings, Part III 38},
  pages={126--152},
  year={2018},
  organization={Springer}
}

@article{ALY23,
  title={Pseudorandom Strings from Pseudorandom Quantum States},
  author={Ananth, Prabhanjan and Lin, Yao-Ting and Yuen, Henry},
  journal={arXiv preprint arXiv:2306.05613},
  year={2023}
}

@misc{BBO+24,
      title={Signatures From Pseudorandom States via $\bot$-PRFs}, 
      author={Mohammed Barhoush and Amit Behera and Lior Ozer and Louis Salvail and Or Sattath},
      year={2024},
      eprint={2311.00847},
      archivePrefix={arXiv},
      primaryClass={cs.CR},
      url={https://arxiv.org/abs/2311.00847}, 
}

@book{M19,
  title={The random matrix theory of the classical compact groups},
  author={Meckes, Elizabeth S},
  volume={218},
  year={2019},
  publisher={Cambridge University Press}
}

@misc{BHM+25,
      author = {Samuel Bouaziz--Ermann and Minki Hhan and Garazi Muguruza and Quoc-Huy Vu},
      title = {On Limits on the Provable Consequences of Quantum Pseudorandomness},
      howpublished = {Cryptology {ePrint} Archive, Paper 2025/1863},
      year = {2025},
      url = {https://eprint.iacr.org/2025/1863}
}
